\documentclass{epl}

\usepackage{graphicx,epsf,epsfig}
\usepackage[T1]{fontenc}
\usepackage{times}

\newcommand{\bq}{{\bf q}}
\newcommand{\bk}{{\bf k}}

\newcommand{\bR}{{\bf R}}
\newcommand{\bz}{{\bf z}}
\newcommand{\etab}{\mbox{\boldmath $\eta $}}
\newcommand{\nustar}{\nu^*}
\newcommand{\nubar}{\bar{\nu}}
\newcommand{\rhobar}{\bar{\rho}}
\newcommand{\chibar}{\bar{\chi}}
\newcommand{\Hhat}{\hat{H}}
\newcommand{\Htilde}{\tilde{H}}
\newcommand{\vtilde}{\tilde{v}}
\newcommand{\ptilde}{\tilde{p}}
\newcommand{\stilde}{\tilde{s}}
\newcommand{\CtwoF}{C$^2$F}
\newcommand{\CjF}{C$^j$F}
\newcommand{\ie}{{\sl i.e. }}
\newcommand{\rhobarbar}{\bar{\bar{\rho}}}
\newcommand{\sbarbar}{\bar{\bar{s}}}

\title{On the self-similarity in quantum Hall systems}
\shorttitle{Self-Similarity in QH Systems}
\author{M.\ O.\ Goerbig\inst{1,2}, P.\ Lederer\inst{2}, and  
C.\ Morais\ Smith\inst{1,3}}
\shortauthor{M. O. Goerbig \etal}

\institute{
\inst{1} D\'epartement de Physique, Universit\'e de Fribourg, P\'erolles,  
CH-1700 Fribourg, Switzerland.\\
\inst{2} Laboratoire de Physique des Solides (associ\'e au CNRS), B\^at.\,510, 
Universit\'e Paris-Sud, F-91405 Orsay cedex, France.\\
\inst{3} Institute for Theoretical Physics, University of Utrecht, 
Leuvenlaan 4, 3584 CE Utrecht, The Netherlands.}

\pacs{73.43.-f}{Quantum Hall effects}
\pacs{73.43.Cd}{Theory and Modelling}
\pacs{71.10.Pm}{Fermions in reduced dimensions}

\begin{document}

\maketitle

\begin{abstract}
The Hall-resistance curve of a two-dimensional electron system in the presence 
of a strong perpendicular magnetic field is an example of self-similarity. It 
reveals plateaus at low temperatures and has a fractal structure. We show that 
this fractal structure emerges naturally in the Hamiltonian formulation of 
composite fermions. After a set of transformations on the electronic model, we 
show that the model, which describes interacting composite fermions in a 
partially filled energy level, is self-similar. This mathematical property 
allows for the construction of a basis of higher generations of composite 
fermions. The collective-excitation dispersion of the recently observed $4/11$
fractional-quantum-Hall state is discussed within the present formalism.
\end{abstract}

\begin{figure}
\centering
\epsfysize+6.5cm
\epsffile{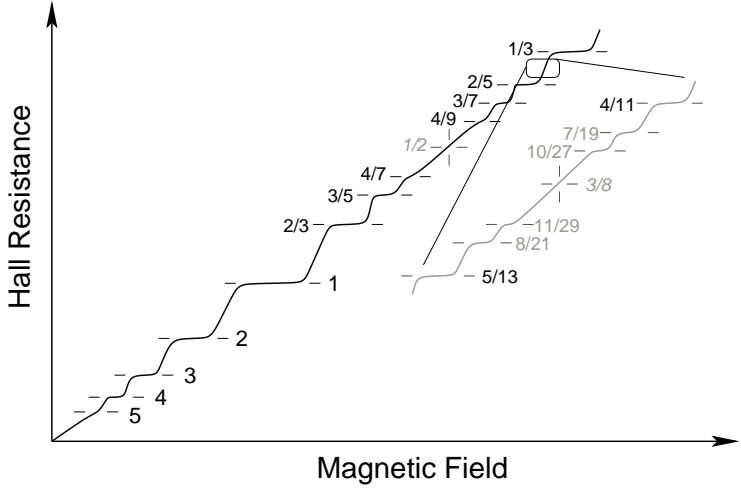}
\caption{Self-similarity of the Hall-resistance curve (c.f. Refs. 2 and 10). 
The plateaus in 
the Hall resistance are due to gapped ground states at certain values of the 
filling factor ($\nu=n$, with integral $n$, for the IQHE, and $\nu=p/(2ps+1)$,
with integral $s$ and $p$, for the FQHE described as an IQHE of CFs). The gray
curve on the right is a zoom on the encircled region of the resistance curve, 
which shows possible states of a second generation of CFs. Only the black 
states ($4/11$ and $5/13$) have been observed in the experiments by Pan 
{\sl et al.} 
\cite{pan}.}
\label{fig01}
\end{figure}

A branch of a snow flake observed under a microscope appears to be almost 
the same as 
the macroscopic flake. This is one of the most prominent examples of 
self-similarity and fractal structure in nature \cite{mandelbrot}. Another 
example of this phenomenon is the curve of the transverse Hall resistance of 
a two-dimensional (2D) electron system in the presence of a strong 
perpendicular magnetic field \cite{mani}: at low temperatures, the primary 
sequence of plateaus in the Hall resistance, which occur at specific values of 
the magnetic field, repeats itself qualitatively in a different field range 
(see Fig.\,\ref{fig01}). The phenomenon of plateau formation in the Hall 
resistance is known as the quantum Hall effect. One distinguishes the integral 
and the fractional quantum Hall effect (IQHE and FQHE). The IQHE 
\cite{klitzing} is found at lower magnetic fields $B$ and 
is a manifestation of the quantisation of the electron energy, with
charge $-e$, into equidistant energy levels, called Landau levels (LLs).
The level degeneracy is characterised by the 
flux density $n_B=eB/h$, and the filling of the levels is given by 
$\nu=n_{el}/n_B$, where $n_{el}$ is the electron density. When $n$ LLs are 
completely filled ($\nu=n$), one finds the IQHE: when the magnetic field
is lowered, the number of states per level is reduced, and some electrons are 
therefore promoted to a higher LL. These electrons become localised due to 
residual impurities in the sample and thus do not contribute to the electrical 
transport. As a consequence, the Hall resistance is not sensitive to the 
change in the magnetic field and remains at its original quantised value 
$R_H=h/e^2n$. This gives rise to a plateau when the Hall resistance is 
plotted as a function of $B$. 

\begin{figure}
\centering
\epsfysize+4.5cm
\epsffile{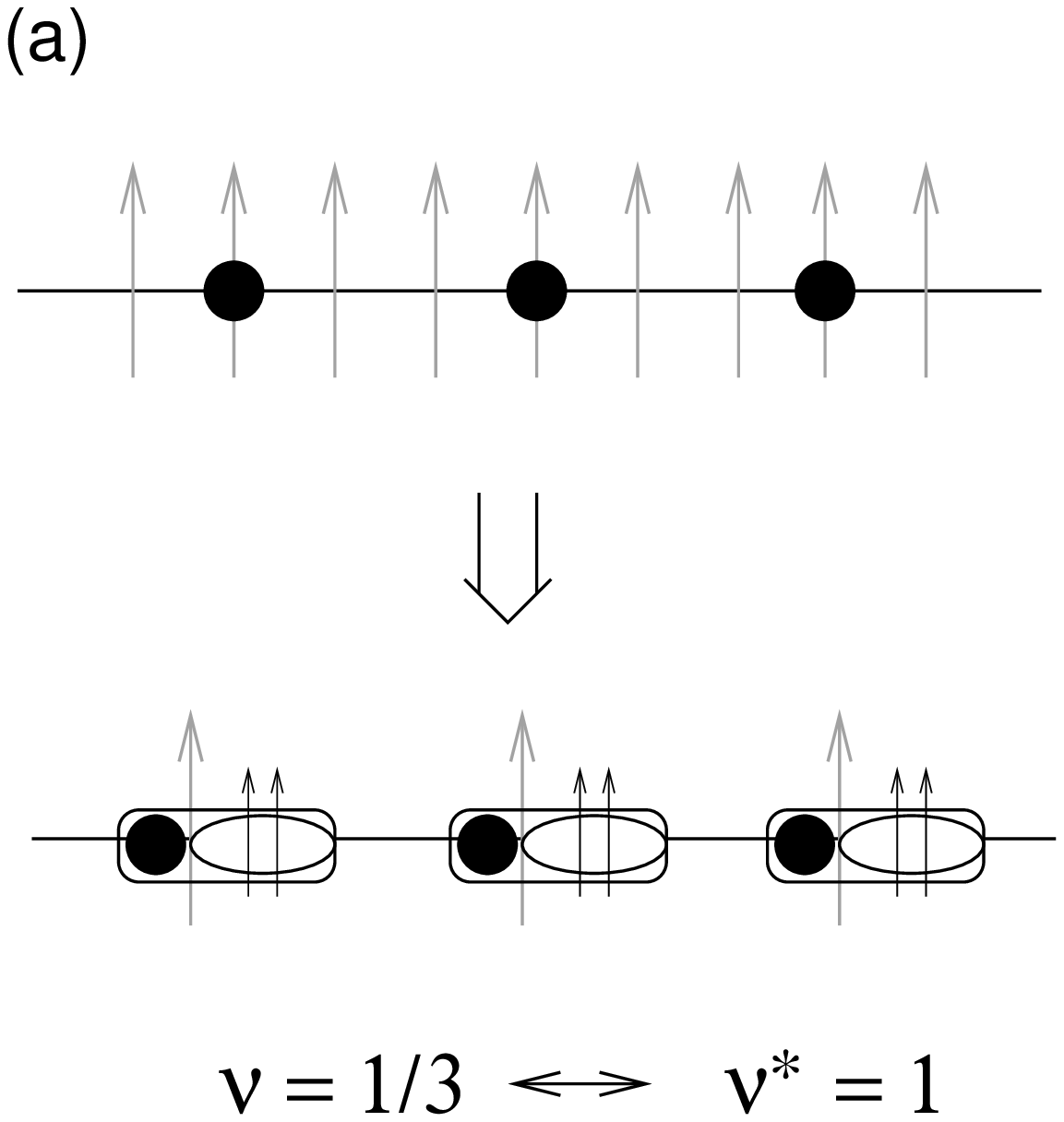}
\epsfysize+4.5cm
\epsffile{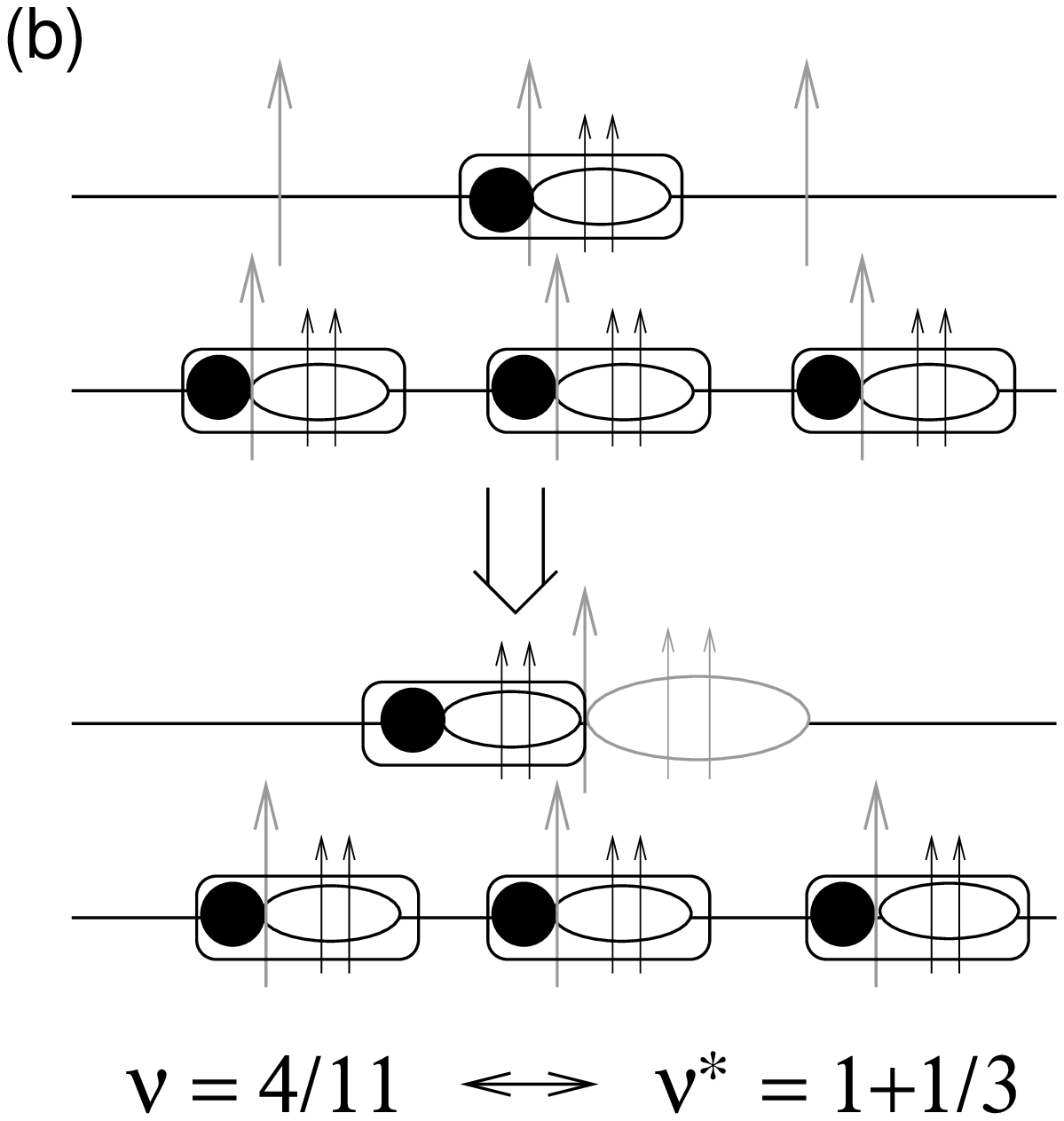}
\caption{Composite fermions. (a) the FQHE at $\nu=1/3$ (top panel) may be 
understood as an IQHE of CFs: each electron (black circle) is bound to a 
vortex-like excitation (open black ellipses) carrying two flux quanta (small 
black arrows). One flux quantum per CF remains ``free'' (big gray arrows) so 
that the effective CF filling factor is $\nustar=1$. (b) sketch of the 4/11 
state if interpreted in terms of a second generation of CFs at 
$\nustar=1+1/3$: whereas the CFs carrying two flux quanta in the lowest CF-LL 
are inert, each CF binds to a pseudovortex (open gray ellipse) with two flux 
quanta (small gray arrows) so that the total number of flux carried by the 
\CtwoF~is four.}
\label{fig02}
\end{figure}

The FQHE is found at fractional values of the filling factor, {\sl i.e.} when 
a LL - mostly the lowest - is only partially filled \cite{stormer}. 
This effect is due to the Coulomb 
interaction between the electrons, which lifts the original degeneracy of the 
LLs. Laughlin explained the existence of gapped ground states at $\nu=1/(2s+1)$
with the help of trial wave functions \cite{laughlin}. Away 
from these filling factors, fractionally charged quasi-particles or 
quasi-holes with a finite energy are excited. In analogy with the IQHE, these 
quasi-particles/-holes become localised, and this leads to the formation of 
plateaus in the Hall resistance around $\nu=1/(2s+1)$. The observation of the 
FQHE at other filling factors such as $\nu=2/5,3/7,4/9,...$, however, required 
new concepts beyond Laughlin's original theory \cite{haldane,halperin}.
Another scheme was put forward by Jain,
who introduced the concept of composite fermions (CFs) \cite{jain}: in order 
to minimise the Coulomb repulsion, an electron binds to a vortex-like 
excitation of the electron liquid carrying $2s$ flux quanta. This bound 
state is exactly the CF [Fig.\,\ref{fig02}(a)]. Due to their renormalised 
charge $e^*=e/(2ps+1)$, the CFs experience a reduced coupling to the magnetic 
field and, in a naive picture, form LLs themselves. The filling $\nu^*$ 
of these CF-LLs is related to the electronic filling factor $\nu$ by
\begin{equation}
\label{equ000}
\nu=\frac{\nustar}{2s\nustar+1}.
\end{equation}
The FQHE at $\nu=p/(2ps+1)$ may be understood as an IQHE of these CFs when 
$p$ CF-LL are completely filled ($\nustar=p$). This theory classifies the most 
prominent FQHE states. 

Recently, Pan \etal~have observed a FQHE at $\nu=4/11$ \cite{pan}, which 
cannot be described in the framework of a theory of non-interacting CFs
\cite{jain}. It corresponds to a CF filling factor 
$\nustar=1+1/3$ and has therefore been interpreted as a FQHE of CFs. It is
natural to generalise the CF picture to higher generations
\cite{smet,goerbig04,lopezfradkin}: whereas the CFs in 
the lowest level are treated as inert, CFs in the partially filled first 
excited CF-LL bind to vortex-like excitations of the CF liquid, and this bound 
state may be viewed as a CF of the second generation [\CtwoF, see 
Fig.\,\ref{fig02}(b)]. An IQHE of \CtwoF s and even higher generations would
lead to the self-similarity of the Hall-resistance curve, pointed out  by Mani
and v.\ Klitzing \cite{mani}.

Here, we provide a mathematical framework for the understanding of the 
self-similarity and the fractal structure of the Hall curve based on the 
Hamiltonian theory of the FQHE, proposed by Murthy and Shankar \cite{MS}. The 
self-similarity of the FQHE is revealed by a reproduction of the original 
quantum-mechanical model, which describes the low-energy degrees of freedom,
after a set of transformations. Because inter-LL excitations belong to the 
high-energy part of the spectrum, which is neglected in the model, 
the kinetic energy is an unimportant constant, and the model of 
electrons restricted to a single LL is given only by the interaction 
Hamiltonian 
\begin{equation}
\label{equ001}
\Hhat=\frac{1}{2}\sum_{\bq}v_n(q)\rhobar(-\bq)\rhobar(\bq),
\end{equation}
where $v_n(q)=(2\pi e^2/\epsilon q)[L_n(q^2l_B^2/2)\exp(-q^2l_B^2/4)]^2$ 
is the effective Coulomb repulsion in the $n$-th LL, in terms of Laguerre 
polynomials $L_n(x)$.
The density operators $\rhobar(\bq)$, {\sl projected} to
a single LL, satisfy the algebra \cite{GMP}
\begin{equation}
\label{equ002}
[\rhobar(\bq),\rhobar(\bk)]=2i\sin\left(\frac{(\bq\times\bk)_zl_B^2}{2}\right)
\rhobar(\bq+\bk).
\end{equation}
The Hamiltonian (\ref{equ001}), together with the commutation relations 
(\ref{equ002}), defines the full quantum mechanical model. The projected 
density operators $\rhobar(\bq)$ may be interpreted as a density of fermions, 
the degrees of freedom of which are described only by the center of their 
cyclotron motion $\bR_j^{(e)}=(X_j,Y_j)$, called the {\sl guiding center} of 
the $j$-th particle,
$$
\rhobar(\bq)=\sum_j e^{-i\bq \cdot\bR_j^{(e)}}=\sum_{m,m'}G_{m,m'}(\bq l_B)
a_m^{\dagger}a_{m'},
$$
where $a_m^{\dagger}$ and $a_m$ are the electron creation and annihilation 
operators in the partially filled LL, respectively. The matrix elements are 
given in 
terms of associated Laguerre polynomials $L_n^m(x)$ (for $m\geq m'$),
\begin{equation}
G_{m,m'}(\bz)=\sqrt{\frac{m'!}{m!}}\left(-\frac{iz}{\sqrt{2}}\right)^{m-m'}
L_{m'}^{m-m'}\left(\frac{|z|^2}{2}\right)e^{-|z|^2/4},
\label{equ002bis}
\end{equation}
with the complex variable $z=x+iy$. The unusual commutation relations 
(\ref{equ002}) arise from the non-commutativity of the components of these 
guiding centers, $[X_k,Y_l]=-il_B^2\delta_{k,l}$,
which is precisely the origin of the fact that each 
quantum-mechanical state occupies a minimal surface $\sigma=2\pi l_B^2=1/n_B$, 
threaded by one flux quantum. The representation of the density operators in 
the CF basis is obtained in two steps. First, one formally introduces the 
density of the vortex-like excitations (``pseudo-vortices''), carrying $2s$ 
flux quanta, $\chibar(\bq)=\sum_j \exp\left(-i\bq \cdot\bR_j^{(v)}\right)$, 
where 
$\bR_j^{(v)}$ is now the guiding center of the $j$-th pseudo-vortex. Because 
the pseudo-vortex has a charge $-c^2=-2ps/(2ps+1)$, in units of the electron 
charge \cite{MS}, its guiding-center components satisfy 
$[X_k^{(v)},Y_l^{(v)}]=i(l_B^2/c^2)\delta_{k,l}$, and this leads to the same
commutation relations for the pseudo-vortex density as for the electron 
density if one replaces $l_B^2\rightarrow -l_B^2/c^2$ in expression 
(\ref{equ002}). This algebraic structure has first been investigated by 
Pasquier and Haldane, who treated a system of {\sl bosons} at $\nu=1$ 
\cite{PH}. The pseudo-vortex artificially introduces new degrees of 
freedom, which would have to be considered in form of a constraint,
$\chibar(\bq)|\psi\rangle=0$, for any physical state $|\psi\rangle$. It must
be taken into account in a conserving approximation if one intends to describe
the physical properties of a {\sl compressible} state at $\nu=1/2$ 
\cite{read2,halperinstern}.

In an approximation, which is valid if the ground state is separated 
from excited states
by an energy gap given by the CF-LL separation \cite{MS}, one may neglect the
constraint and construct the CF density as a superposition of the electronic 
and pseudo-vortex densities, 
$$\rhobar_{CF}(\bq)=\rhobar(\bq)-c^2\chibar(\bq).$$
In a second step, one performs a variable change to introduce the CF guiding 
center as a weighted sum of the electron and pseudo-vortex guiding centers 
\cite{MS},
\begin{equation}
\label{equ003}
\bR^{CF}=\frac{\bR^{(e)}-c^2\bR^{(v)}}{1-c^2},\qquad
\etab^{CF}=\frac{c}{1-c^2}\left(\bR^{(e)}-\bR^{(v)}\right).
\end{equation}
The operator $\etab^{CF}$ may be interpreted as a CF cyclotron coordinate, 
responsible for the formation of the CF-LLs.
The degeneracy of any CF-LL is obtained from the minimal surface 
occupied by each state $\sigma^*=2\pi l_B^{*2}$ due to the commutation 
relations of the CF guiding-center components, $[X^{CF},Y^{CF}]=-il_B^{*2}$, 
where $l_B^*=l_B/\sqrt{1-c^2}$ is the CF magnetic length. This yields a 
reduced flux density $n_B^*=1/2\pi l_B^{*2}$, and one obtains the IQHE of CFs
if $p$ CF-LL are completely filled ($\nustar=n_{el}/n_B^*=p$).
 
In analogy with the electronic case, the excitations of lowest energy are 
those within the same CF-LL $p$, and they are only present if this level is 
partially filled, \ie if $p<\nustar<p+1$. The CFs in the completely filled 
lower levels may be considered as inert. Formally this means that one 
restricts the CF-density operator to the $p$-th level, 
$\langle \rhobar_{CF}(\bq)\rangle_{p}=F_{CF}^p(q) \rhobarbar (\bq)$, with the 
projected CF-density operator
$$\rhobarbar (\bq)=\sum_j e^{-i\bq \cdot\bR_j^{CF}}=
\sum_{m,m'}G_{m,m'}(\bq l_B^*)b_m^{\dagger}b_{m'},
$$
where $b_m^{\dagger}$ and $b_m$ are the CF creation and annihilation operators 
in the $p$-th CF-LL. The matrix elements, given by Eq.\,(\ref{equ002bis}), are 
the same as for the electronic case in terms of the renormalised magnetic 
length $l_B^*$. For the projected CF density operators, one obtains exactly
the same algebra as for the projected electron densities (\ref{equ002}) if 
one replaces $l_B\rightarrow l_B^*$,
\begin{equation}
\label{equ010}
[\rhobarbar(\bq),\rhobarbar(\bk)]=
2i\sin\left(\frac{(\bq\times\bk)_zl_B^{*2}}{2}\right)\rhobarbar(\bq+\bk).
\end{equation}
The new model Hamiltonian for restricted CFs is 
\begin{equation}
\label{equ011}
\Htilde=\frac{1}{2}\sum_{\bq}\vtilde(q)\rhobarbar(-\bq)\rhobarbar(\bq),
\end{equation}
where the CF form factor of the $p$-th CF-LL,
$$
F_{CF}^p(q)= e^{-q^2l_B^{*2}c^2/4}\left[L_p\left(\frac{q^2l_B^{*2}c^2}{2}
\right)-c^2e^{-q^2l_B^2/2c^2}L_p\left(\frac{q^2l_B^{*2}}{2c^2}\right)\right],
$$
has been absorbed into an effective CF interaction potential 
$\vtilde(q)=v_0(q)[F_{CF}^p(q)]^2$, for the case of the lowest electronic 
LL $n=0$. The fact that one obtains a model with the
same structure after restriction of the CF dynamics to a fixed CF-LL is
precisely a mathematical manifestation of self-similarity. One 
simply has to take into account a renormalisation of the magnetic length and 
replace the original interaction potential by the effective CF potential.
In analogy with the CF basis described above, on may find a 
representation in terms of \CtwoF s: after the introduction of a new 
pseudo-vortex density operator, one performs a variable change such as given 
by Eq.\,(\ref{equ003}). These 
\CtwoF -states may in principle be found at CF fillings 
$\nustar=p+\ptilde/(2\stilde\ptilde+1)$, where each \CtwoF~carries $2\stilde$ 
flux quanta, and $\ptilde$ denotes the number of completely filled \CtwoF-LLs. 
In this approach, the $4/11$ state corresponds to $s=p=\stilde=\ptilde=1$. 

The iteration of this projection and the construction of such a new basis 
provides a possible framework for the description of higher generations of CFs.
This approach yields a recursion 
formula for the filling factor $\nu_j$ of the $j$-th 
generation of CFs (\CjF s),
\begin{equation}
\label{equ012}
\nu_{j}=p_{j}+\frac{\nu_{j+1}}{2s_{j+1}\nu_{j+1}+1},
\end{equation}
where $s_{j+1}$ denotes the number of flux pairs carried by the pseudo-vortex 
in the C$^{j+1}$F, and $p_{j}$ is the number of completely filled \CjF-LLs. 
The IQHE of \CjF~is obtained for $\nu_j=p_j$, with integral $p_j$. Formally, 
the electronic case corresponds to $j=0$, and thus $\nu_0=\nu$ and 
$\nu_1=\nustar$. Eq.\,(\ref{equ012}) is a generalisation of the relation 
(\ref{equ000}) between the electronic and the CF filling factors.

Note that the mathematical self-similarity on the level of the model structure
does not {\sl guarantee} a self-similarity between states.
The existence of higher-generation CFs depends on the precise form
of the CF interaction potential, which is {\sl different} from the electronic
one. Also the existence of first-generation CF states
depends on the form of the effective interaction potential, which varies with 
the LL index although the structure of the model remains the same. Whereas in 
the two lowest LLs a $1/3$ FQHE state has been observed, such a state has a 
higher energy than a two-electron bubble crystal in the second excited LL 
\cite{fogler,goerbig03}. The present model provides a framework for 
studying the competing phases in the case of \CtwoF s. Detailed energy 
calculations, performed directly in the thermodynamic limit, indicate that 
several \CtwoF~states are stable \cite{goerbig08}, in contrast to 
numerical-diagonalisation studies in Jain's wave-function approach. The latter 
method is restricted to the study of small systems and yields an 
alternation between compressible and incompressible states at $\nu=4/11$ as
a function of the diagonalised number of CFs \cite{mandaljain}. The 
extrapolation of these ambiguous results to the thermodynamic limit has misled
the authors in the interpretation of this state \cite{mandaljain}.

\begin{figure}
\centering
\epsfysize+6cm
\epsffile{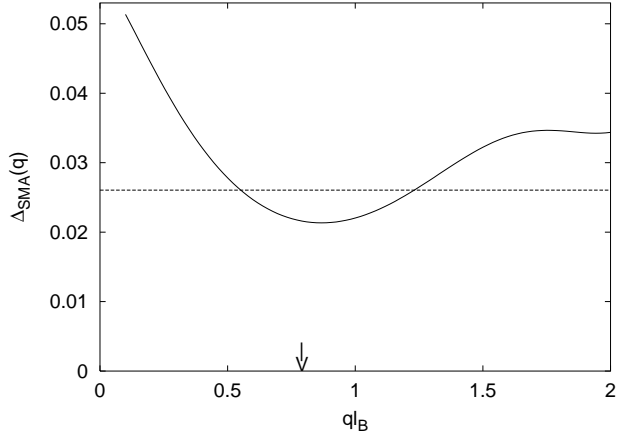}
\caption{Dispersion of collective excitations at $\nu=4/11$,
  calculated in the SMA. The arrow indicates the modulus of the
  reciprocal lattice vector of a CF Wigner crystal.
  The dashed line represents the value of the activation
  gap of the \CtwoF~state, obtained in the Hamiltonian theory.}
\label{fig03}
\end{figure}

At present, the only observed FQHE state, which may be interpreted in terms of
spin-polarised \CtwoF s, is the $4/11$ state \cite{pan}. Our formalism
allows one to study various physical quantities of this state, such as
its activation gap \cite{goerbig04} or its energy with respect to
competing CF-solid phases \cite{goerbig08}. Here, we investigate the
dispersion of its collective excitations in the single-mode
approximation (SMA), which has first been applied to the Laughlin states in the
lowest LL \cite{GMP}. The excited state at $\nu=4/11$ may be obtained
in the SMA by application of the projected CF-density operator
$\rhobarbar(\bq)$ on the Laughlin state of CFs $|\Omega\rangle$, 
$|\psi_{\bq}\rangle=\rhobarbar(\bq)|\Omega\rangle$. Its energy
$\Delta_{SMA}(q)$ may be
expressed in terms of the projected static structure factor \cite{goerbig01},
$$\sbarbar(q)=\frac{\langle\Omega|\rhobarbar(-\bq)\rhobarbar(\bq)|
\Omega\rangle}{N_{el}}=(1-\nubar^*)+4\nubar^*\sum_{m=0}^{\infty}
c_{2m+1}L_{2m+1}\left(q^2l_B^{*2}\right)e^{-q^2l_B^{*2}/2},$$
where $\nubar^*=1/3$ is the filling of the CF-LL $p=1$, and the expansion
coefficients $c_{2m+1}$ characterise the Laughlin state
$|\Omega\rangle$. The energy of the collective excitations,
\begin{equation}
\label{equ013}
\Delta_{SMA}(q)=2\sum_{\bk}\left[\vtilde(|\bk-\bq|)-
\vtilde(k)\right]\sin^2\left(\frac{(k_xq_y-k_yq_x)l_B^{*2}}{2}\right)
\frac{\sbarbar(k)}{\sbarbar(q)},
\end{equation}
is shown in Fig. \ref{fig03}. The position of the {\sl roton-minimum} 
$ql_B\simeq0.85$ matches well the modulus of the reciprocal
lattice vector of a CF Wigner crystal $ql_B=\sqrt{\pi/3^{3/2}}\simeq0.78$,
which one obtains from $\nubar^*=2\pi l_B^{*2}/A_{pc}$, with the area of
the primitive cell $A_{pc}$ of a triangular lattice. In analogy with 
the
interpretation of the magneto-roton of the electronic Laughlin states
\cite{GMP}, one may view it as a tendency of the liquid state to
crystallise. The dashed line in Fig. \ref{fig03} represents the energy
of the activation gap, calculated in the Hamiltonian theory
\cite{goerbig04}. Note that in contrast to Ref. \cite{goerbig04},
screening effects due to inter-CF-LL excitations, as well as finite-width 
effects, have been neglected here. In principle, the activation
gap should coincide with the dispersion at large wave
vectors, but it is known that the SMA yields less reliable
results at large wave vectors \cite{GMP}. However, the fact that the
dispersion has a lower energy than the activation gap at the
roton-minimum indicates that the SMA is accurate at intermediate values of $q$.

In conclusion, we have investigated the self-similarity of the
quantum Hall effect in the framework of the Hamiltonian theory
\cite{MS}. A change from the
electronic to the CF basis - or more generally from C$^{j-1}$Fs to
\CjF s - and the restriction of the particle dynamics to a fixed level
yield the same model as the original one if one rescales the magnetic
length and takes into account a modification of the interaction
potential. Both the Hamiltonian [Eqs.\,(\ref{equ001}) and
(\ref{equ011})] and the commutation relations [Eqs.\,(\ref{equ002})
and (\ref{equ010})] are reproduced in this projection scheme. 
However, it is not clear whether higher generations of CFs 
than \CtwoF s are stable even in extremely pure samples; in this case, 
the CF-LL separation becomes extremely tiny, and the restriction 
to one level may therefore be a rather poor approximation because residual CF 
interactions are expected to lead to a strong level mixing. As an
illustration of the relevance of the interacting-CF model, we have calculated 
the collective-excitation dispersion at $\nu=4/11$ in the SMA and
compared the results to the \CtwoF~activation gaps \cite{goerbig04}.
These results may shed some light on the similarities 
between electrons and (higher-generation) CFs, which appear to be essential 
for the understanding of QHE systems.

\vspace{-0.3cm}
\acknowledgments
\vspace{-0.3cm}
We thank R.\ Morf for fruitful discussions and R.\ G.\ Mani for pointing out 
to us Ref. \cite{mani}, which stimulated the present studies. This work was 
supported by the Swiss National Foundation for Scientific Research under 
grant No.~620-62868.00.

\end{document}